\documentclass[pra,twocolumn,showpacs,preprintnumbers,amsmath,amssymb]{revtex4}
\usepackage{graphicx}
\usepackage{epsfig}
\usepackage{bm}
\newcommand{\be}{\begin{equation}}
\newcommand{\ee}{\end{equation}}
\newcommand{\bea}{\begin{eqnarray}}
\newcommand{\eea}{\end{eqnarray}}

\newcommand{\ket}[1]{ \mid#1\rangle}
\newcommand{\bra}[1]{\langle#1\mid}
\newcommand{\mat}[1]{\mathcal{#1}}
\newcommand{\abs}[1]{\left|#1\right|}
\newcommand{\laj}{\lambda_j}

\newcommand{\roj}{\abs{\rho_j}^2}
\newcommand{\sij}{\abs{\sigma_j}^2}
\newcommand{\gaj}{\abs{\gamma_j}^2}
\begin{document}
\title{Perfect state transfer in  long-range interacting spin chains}
\author{Giulia Gualdi$^\dagger$, Vojtech Kostak$^\ddagger$, 
        Irene Marzoli$^\dagger$, Paolo Tombesi$^\dagger$}
\affiliation{$^\dagger$Dipartimento di Fisica, Universit\`a degli Studi di 
             Camerino, I-62032 Camerino, Italy \\
             $^\ddagger$Department of Physics, FJFI \v{C}VUT,  B\v{r}ehov\'a 7,
              115 19 Praha, Star\'e M\v{e}sto, Czech Republic}
\date{\today}
\begin{abstract}
We investigate the most general conditions under which a finite ferromagnetic 
long-range interacting spin chain achieves unitary fidelity and the shortest 
transfer time in transmitting an unknown input qubit. 
A deeper insight into system dynamics, allows us to identify an ideal system 
involving sender and receiver only. 
However, this two-spin ideal chain is unpractical due to the rapid decrease of 
the coupling strength with the distance. 
Therefore, we propose an optimization scheme for approaching the ideal 
behaviour, while keeping the interaction strength still reasonably high. 
The procedure  is scalable with the size of the system and straightforward 
to implement.
\end{abstract}
\maketitle
\section{Introduction}
The majority of protocols for quantum communication relies on photons \cite{brau}, 
because of their weak interaction with the environment and of the well-developed optical fiber 
technology. However, it is not always convenient to use photons to exchange information.  In fact, when dealing  with quantum processors, it is not straightforward to convert a stationary qubit into a flying one and viceversa. An alternative scheme is based
on repeated swapping operations which, however, require a carefully designed and controlled sequence of pulses. Therefore it is highly desirable to achieve state transfer by just letting a system evolve. Indeed, this is the case of interacting spin chains which can serve as quantum channels for short or mid-range quantum communication \cite{B, giov, cr,ks,Has}. 
Moreover, trapped particles, such as ions or electrons, not only are suitable to implement a scalable quantum processor \cite{ciar1,cira1}, but are also able to reproduce an effective spin-spin coupling which exhibits dipolar decay \cite{ciar,cira}.

In this paper, we show a simple way to attain perfect state transfer with a finite spin chain  exhibiting the most general  long-range (LR) interaction. Previous theoretical work concentrated mostly on the idealized case of nearest-neighbour interactions,  with few exceptions \cite{kay,afb}, and often even restricted to XY spin chains \cite{plenio, woj}. Our procedure consists in removing sender and receiver nearest-neighbours, thus exploiting a peculiar feature of the LR coupling, and selectively acting on the system eigenvectors.  The resulting quantum channel exhibits optimal performances when compared to the ideal system with sender and receiver only. Moreover, the fidelity becomes practically invariant under system scaling and the transfer time independent of the number of spins.

The paper is organized as follows. 
In Sec.~\ref{conditions} we analytically derive the most general
conditions, under which a finite ferromagnetic system of interacting spins achieves unitary fidelity
in the transmission of an unknown qubit state from a sender to a receiver site.
We then apply our analysis to the case of a long-range interacting spin chain (Sec.~\ref{LR_chain}) and
propose our strategy to optimize the system performances in terms of fidelity and transfer time
(Sec.~\ref{DH_chain}). 
Finally, we summarize our results and discuss possible experimental implementations for our scheme
(Sec.~\ref{conclusions}).
\section{Conditions for perfect state transfer}
\label{conditions}
Given a generic ferromagnetic system of $N$ spins, we assume that its Hamiltonian $H$ preserves 
the total magnetization $M\equiv\sum_{i=1}^N S^z_i$ such that $[H,M]=0$, with $S_i^z$ being the $z$ component of the total spin operator $\bm S_i$ of the $i$-th spin.
 This implies that the initial state evolves only into states with the same number of excitations.
As a basis for the $N\times N$ single excitation subspace, which is relevant for the qubit transfer, we adopt the set of vectors $\bm{\ket j}=\bigotimes_{k=1;k\neq j}^N\ket 0_k \otimes\ket 1_j$, where $j$ indicates the site where the spin has been flipped from 0 to 1 \cite{bethe}. Moreover, 
we denote with $\bm{\ket 0}=\bigotimes_{k=1}^N\ket{0}_k$ the ground state of the system, with all the  spins facing down parallel to the external magnetic field.
Our task is to transmit an unknown input state from a sender site $s$ to a receiver site $r$. The performance of the quantum channel is measured by the fidelity \cite{B}
\be F(t)=\frac{\abs{f(t)}^2}{6}+\frac{\abs{f(t)}}{3}+\frac{1}{2},\label{Fid}\ee
where \be f(t)=\bm{\bra{r}}e^{-iHt}\bm{\ket{s}}\label{prop}\ee is the propagator of the excitation from sender to receiver with $\hbar$ equal to 1. From Eq. (\ref{Fid}), it is clear that the fidelity reaches its maximum value, one, iff $\abs{f(t)}^2=1$. Hence, we investigate under which conditions $\abs{f(t)}^2$ takes on the unitary value. Given the set of eigenvectors $\{\ket\laj\}$ with eigenvalues $\{E_j\}$, such that $H\ket\laj=E_j\ket\laj$, we can expand $\abs{f(t)}^2$ in terms of the system eigenstates\be
\abs{f(t)}^2=\mid\sum_{j=1}^Ne^{-iE_jt}\bm{\bra r}\laj\rangle\langle\laj \bm{\ket s}\mid^2,\label{fsqua}\ee
in order to obtain\be \abs{f(t)}^2=f_m+f_t,\label{fsqua2}\ee with \bea
f_m&\equiv&\sum_{j=1}^N\abs{\sigma_j}^2\abs{\rho_j}^2,\label{fm}\\
f_t&\equiv&2\sum_{k<l}^N\abs{\sigma_k}\abs{\sigma_l}\abs{\rho_k}\abs{\rho_l}\cos(\Delta_{k,l}t+\xi_{k,l}),\label{ft}\eea
where $\abs{\sigma_j}e^{\phi_j}$ ($\abs{\rho_j}e^{\psi_j}$) is the projection of the eigenvector $\ket\laj$ on the initial (final) state, when the excitation is located at site $s$ ($r$) and $\xi_{k,l}\equiv\phi_k-\phi_l-\psi_k+\psi_l$.  From Eq. (\ref{fsqua2}) we note that $\abs{f(t)}^2$ consists  of two terms. The first one, $f_m$ is time-independent, whereas the second one, $f_t$, oscillates with frequencies $\Delta_{k,l}\equiv E_k-E_l$. Besides very specific cases of mirror-periodic systems, or  locally approximable as such \cite{cr,ks}, these frequencies are uncorrelated. Hence, in a generic case, the time-dependent term, $f_t$, oscillates {\it almost} symmetrically around its average value which is approximately zero. Since $\abs{f(t)}^2\in[0,1]$, the allowed minimum value for $f_t$ is $-f_m$. Therefore, up to fast oscillations with very narrow peaks, it seems reasonable to approximate the upper bound of $f_t$ with $f_m$. 
This implies that $2f_m$ is an accurate estimate for the maximum value of $\abs{f(t)}^2$. Hence, we investigate when $f_m$ reaches its maximum value. This search is bounded by the normalization constrains
\bea
&&\sum_{j=1}^N\abs{\sigma_j}^2=\sum_{j=1}^N\abs{\rho_j}^2\equiv 1,\label{sigroj}\\
&&\abs{\sigma_j}^2+\abs{\rho_j}^2+\abs{\gamma_j}^2\equiv1,\;\forall j.\label{gaj}\eea
We interpret $\abs{\gamma_j}^2$ as the overlap of the $j_{th}$ eigenvector with all the basis states besides $\bm{\ket s}$ and $\bm{\ket r}$ \be
\abs{\gamma_j}^2=\sum_{i\neq(s,r)}\mid\langle\laj\bm{\ket i}\mid^2.\label{gamma}\ee
Keeping $\gaj$ fixed, we look for the extremal points of $f_m$. Necessary condition for extremes is the  spatial symmetry between the projections of the eigenvectors on initial and final states. Thus, the local maximum is reached for\be \sij=\roj=\frac{1-\gaj}{2}.\label{sym}\ee
Indeed, given a number $N$ of spins, the fidelity is maximized iff sender and receiver are located symmetrically with respect to the midpoint of their joining axis. Now, the absolute maximum of $f_m$ depends on $\gaj$. Given the constrain $\sum_{j=1}^N\gaj=N-2$, the global extremal point $f_m=1/N$ is reached when $\gaj=(N-2)/N$, for each $j$.
Due to its decreasing monotonicity as a function of $N$, $f_m$ reaches its absolute maximum on the lower border of its domain, i.e. for $N=2$. Thus, to achieve perfect state transfer when $N>2$, only two eigenvectors $\ket{\lambda^{id}_{\pm}}$ must have finite projections on initial and final states and zero projections on the other basis states\bea
&&\sij=\roj=\frac{1}{2},\;\;\gaj=0,\quad\;\;j=+,-\label{globmax1}\\
&&\sij=\roj=0,\;\;\gaj=1,\quad\;\;\mbox{otherwise}.\label{globmax2}\eea
The global maximum conditions in Eqs. (\ref{globmax1}) and (\ref{globmax2}) state that unitary fidelity can be attained iff the Hilbert space of the system is the direct sum of two disjoint subspaces, i.e. $\mat H^{N\times N}_{system}\equiv\mat H^{2\times 2}_{s,r}\oplus\mat H^{(N-2)\times(N-2)}_{channel}$, pertaining, respectively, to the sender-receiver pair and to the rest of the chain. The mixing between these two subspaces is measured by the quantity $\abs{\gamma_\pm}^2$, which, in fact, must be zero  to satisfy the maximum conditions.  In other words, only two spins, at the sender and receiver sites, and two eigenvectors, the symmetric and antisymmetric combination of the initial and final states, must play a role in the communication. Hence, we obtain an ideal ($id$) system, whose eigenvectors are
\be \ket{\lambda^{id}_{\pm}}=\frac{1}{\sqrt2}(\bm{\ket s}\pm\bm{\ket r}). \label{lid}\ee
\section{The long-range interacting spin chain}
\label{LR_chain}
We now focus on linear LR interacting spin chains represented by the most general XYZ Heisenberg Hamiltonian\be
H=\sum_{i,j;i\neq j}^NJ_{i,j}(\bm S_i\cdot\bm S_j-3S^z_iS^z_j)\quad\mbox{with}\; J_{i,j}=\frac{C}{(a\abs{i-j})^{\nu}},\label{Hnu}\ee
where $\nu>0$, $a$ is the fixed inter-spin distance and $C$ is a model depending constant. In particular, the case $\nu=3$ corresponds to the dipolar coupling.
Hence, the energy between nearest neighbours is $\bm{\bra i}H\bm{\ket{i+1}}=C/(2a^\nu)$. We choose the energy, length and time units by setting this last quantity and $a$ equal to unity \cite{afb}. In the ideal case, where the sender-receiver subspace is completely detached from the rest of chain, the eigenenergies corresponding to the eigenvectors in Eq. (\ref{lid}) read\be E^{id}_+=0,\quad E_-^{id}=-\frac{2}{(N-1)^\nu},\label{eid}\ee 
where $N-1$ is the number of length units between sender and receiver.
In this case,  Eq. (\ref{fsqua2}) gives $\abs{f(t)}^2=\sin^2(\mid E_+^{id}-E_-^{id}\mid t/2)$, which leads to perfect state transfer, i.e. unitary fidelity, for \be t_{id}=\frac{\pi}{2}\;(N-1)^{\nu}.\label{tid}\ee As expected, the transfer time, $t_{id}$, increases with the transmission distance according to a power law depending on the specific LR interaction.

\begin{figure}
\includegraphics[width=7cm]{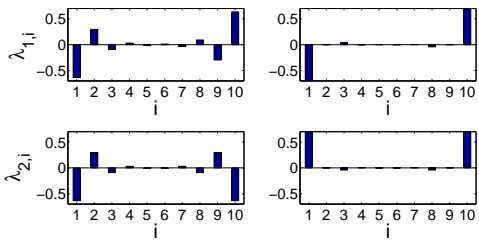}
\caption{\label{eigv10}Numerically calculated components $\lambda_{j,i}\equiv\langle\lambda_j\bm{\ket i}$, with $j=1,2$, of the first two low-lying eigenvectors for a dipolar chain of ten sites. Left column: the complete chain with ten spins. Right column: the DH chain (i.e. the same chain without the spins located at sites 2 and 9).}
\end{figure}
\begin{figure}
\includegraphics[width=6cm]{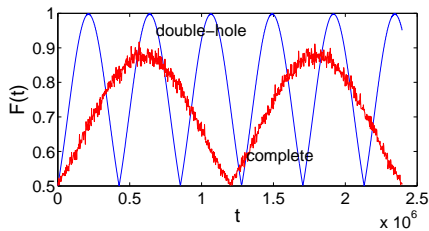}\caption{\label{confid50}(colour online) The  fidelity as a function of time for a transmission distance of $N-1=49$ length units. The red line corresponds to the complete chain, whereas the blue line corresponds to the DH chain.}
\end{figure}
\begin{figure}
\includegraphics[width=6cm]{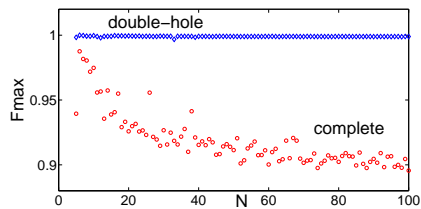}\caption{\label{confid2}(colour online) Maximum values of fidelity as a function of the number of sites $N$ in the complete chain (red circles) and in the DH case (blue diamonds).}
\end{figure}
\begin{figure}
\includegraphics[width=6cm]{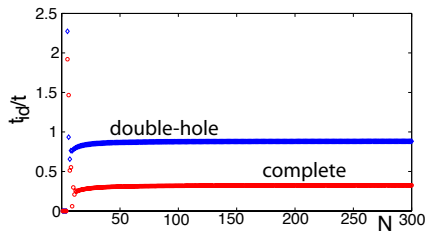}\caption{\label{trtconf}(colour online) The ratio between the ideal transfer time $t_{id}$ and the computed transfer time $t$ as a function of the number of sites $N$ for  the complete chain (red circles) and the DH system (blue diamonds).}
\end{figure}
In a realistic case the eigenvectors tend to have non-zero projections on all the basis states. However, two of them, $\ket{\lambda_1}$ and $\ket{\lambda_2}$, corresponding to the lowest eigenvalues, show a significant  overlap with $\bm{\ket 1}$ and $\bm{\ket N}$. 
  This is due to the  linear topology, where the outermost spins are easier to flip, being less bound to the rest of the chain. Hence, the external sites represent the optimal sender-receiver pair. 
As an example we plot in Fig. \ref{eigv10} (left column) the spatial distribution of the first couple of eigenvectors for a dipolar ($\nu=3$) chain of ten spins.
We note that  the overlap $\abs{\gamma_{j}}^2$ ($j=1,2$), with the other basis states, is non negligible only with $\bm{\ket2},\bm{\ket{N-1}}$.
Thus,  we approximate the system dynamics taking into account only  the two low-lying eigenvectors. Therefore, in analogy to the ideal case,  we can estimate the transfer time of a finite linear chain as \be
t\approx\frac{\pi}{\Delta_{12}},\label{tr}\ee where $\Delta_{12}$ is the energy difference between the two lowest eigenvalues.  Equation (\ref{tr}) is in agreement with the value of the transfer time provided in Ref. \cite{afb} for the dipole-like interaction. However, our result is more general, since it applies to any finite linear ferromagnetic system of interacting spins.
\section{The double-hole chain}
\label{DH_chain}
To reduce the mixing between the sender-receiver subspace and the rest of the chain, we remove the symmetric couple of spins located at the  sites 2 and $N-1$. Therefore, in the Hamiltonian Eq. (\ref{Hnu}), we set to zero the coupling constants $J_{2,i}$ and $J_{N-1,i}$, for each $i$. 
 Sender and receiver are still found at the two ends of the chain, but now their nearest neighbouring sites are  empty. This implies that the coupling strength, between sender and receiver and their new respective nearest-neighbour, is decreased, in the case of the double-hole (DH) chain, by a factor $2^{\nu}$ with respect to the original complete chain. This procedure preserves the overall system symmetry, while it increases the energy separation between the sender-receiver subspace and the rest of the chain. 
Indeed, we observe, right column of Fig. \ref{eigv10}, a more pronounced localization of the two eigenvectors that makes the DH chain more closely resembling the ideal case. 

Let us now characterize  the DH system  performances, as a quantum channel, in terms of fidelity and transfer time. To this end, we compute the fidelity according to Eq. (\ref{Fid}), for a dipole-like ($\nu=3$) interacting spin chain. In Fig. \ref{confid50}, for a given sender-receiver distance,  we compare the performances of the complete chain and of the DH chain. Three major features emerge: $i)$ the DH chain attains unitary fidelity, whereas the complete chain barely reaches 0.9; $ii)$ the DH chain is about three times faster in transferring the qubit state; $iii)$ the DH fidelity is a smooth function of time, well approximated by a sinusoidal behaviour. This dramatic improvement of the chain performances is due to the reduced mixing between the sender-receiver subspace and the rest of the chain, achieved in the DH configuration ($\abs{\gamma_j}^{2}_{DH}\ll\abs{\gamma_j}^2$, with $j=1,2$).  From Fig. \ref{confid2}, it is apparent that the DH system mantains a practically perfect fidelity for at least 100 sites. We did not investigate longer chains just because of computational time restrictions. Not only the DH chain outperforms the full chain, but also the  maximum fidelity is almost insensitive to both the distance and the number of spins between sender and receiver. Moreover given a fixed transmission distance, both fidelity and transfer time become invariant under system rescaling in the limit $\gaj\rightarrow0$, with $j=1,2$. Indeed, as it appears in Fig. \ref{trtconf}, the ratio between the ideal transfer time $t_{id}$ and the DH transfer time  approaches the asymptotical value of $t_{id}/t=0.883$, whereas  for the complete chain this ratio tends to 0.326. Despite the increased distance between sender and receiver and the rest of chain, the message transmission is qualitatively (higher fidelity) and quantitatively (shorter transfer time) enhanced in the DH setup.
 We can regard the spins interposed between sender and receiver as repeaters, whose reflectance is proportional to the overlap of the corresponding basis states with the lowest eigenvectors. Hence, we are led to interpret $\gaj$, with $j=1,2$, as the channel opacity. Indeed for the ideal system, Eq. (\ref{globmax1}), this quantity is zero. Therefore  $t_{id}$, Eq. (\ref{tid}), provides the lowest bound to the transfer time.
 The DH transfer time nearly approaches this minimum, thus marking a great improvement over the complete chain. 
 \begin{figure}
\includegraphics[width=6cm]{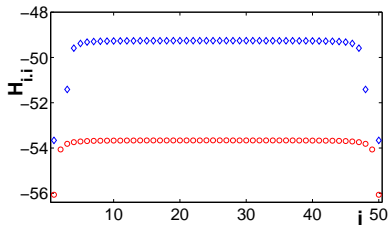}\caption{\label{onsite50} (colour online) Plot of the  diagonal elements $H_{i,i}\equiv\bm{\bra i} H\bm{\ket i}$ of the Hamiltonian Eq. (\ref{Hnu}), as a function of the excitation position for the complete chain (red circles) and the DH chain (blue diamonds).}
\end{figure}
 The transfer time is also a measure of the stability of the system against small perturbations affecting the channel between sender and receiver (variations of the coupling, non-uniform filling, variable inter-spin spacing, etc\ldots). Indeed, the transfer time decreases proportionally to how much the system approximates the ideal two-spin dynamics. This goal is achieved by separating the two lowest eigenvalues from the rest of the spectrum, i.e. by confining the system in a well defined portion of its phase-space, which is energetically expensive to leave. From this point of view, the DH system is stabler than its complete counterpart. 
The diagonal terms  $\bm{\bra i}H\bm{\ket i}$ of the Hamiltonian, Eq. (\ref{Hnu}),
represent the configuration energy when the spin at the $i$-th site has been flipped from 0 to 1. From Fig. \ref{onsite50}, we see that the lowest energy pertains to the configurations where the excitation is localized at the extremes of the chain. Moving from the complete to the DH chain, the energy difference between  these configurations  and the rest of the chain increases, thus improving the excitation confinement to this region. Therefore we expect that perturbations in the channel are less likely to affect the system dynamics and, in this respect, the DH performances are more robust than their complete counterpart. 
\section{Conclusions}
\label{conclusions}
We envisage possible implementations  of this scheme based  on trapped particles, 
such as electrons  or ions. 
Indeed,  these systems realize an effective spin-spin dipole-like coupling with the experimental 
control over the interaction strength \cite{ciar,cira}. 
Microtrap arrays allow for a more  accurate design of the inter-spin distance, whereas, for ions in 
linear Paul traps, one has to devise a strategy to switch-off the interaction between sender-receiver 
and their nearest-neighbours.  
Thanks to the single particle addressability with a laser beam, this could be accomplished by putting 
out of resonance the ions  sitting at sites $(2,N-1)$ with respect to the driving field used to 
estabilish the spin-spin interaction. 
We emphasize that not only the DH chain requires the same technology as the complete chain, 
but also,
due to the negligible overlap of the relevant eigenvectors with the other basis states corresponding to the rest of the chain, its performances are definitely more robust against  experimental defects. Moreover, the smooth time behaviour of the DH  fidelity greatly relaxes the necessary time resolution  for the accomplishment of the communication protocol.

Summarizing, in this paper we have proved that perfect state transfer takes place when 
the sender-receiver subspace is detached from the rest of the chain. 
This condition implies that ideally only two eigenvectors are essential for the communication. 
A similar state transfer protocol has been proposed for antiferromagnetic spin chains \cite{lde1}, 
exploiting their peculiar SU(2) global invariance. 
Therefore it is quite surprising to see that unitary fidelity can be achieved also with ferromagnetic 
systems. 
We have outlined a scalable procedure, without neither additional resources, nor demanding  
pre-engineering or dynamical control of  the couplings. 
In case of dipolar interaction, our numerical estimates show that, given a fixed transmission 
distance, fidelity and transfer time approach the ideal values and, most notably, are invariant under 
system rescaling. 
This procedure can be extended to all  LR interacting systems, simply by adjusting  the number 
of neighbouring spins to be removed, in order to obtain optimal performances.
\acknowledgments
This research was supported by the European Commission through the STREP QUELE, the  IP FET/QIPC SCALA, and the Marie Curie RTN CONQUEST, and by the Italian MUR through PRIN 2005. G. G. is grateful to Dr. G. Di Giuseppe for useful discussions and suggestions.

\end{document}